\documentclass{PoS}

\newcommand{\Slash}[1]{{\ooalign{\hfil/\hfil\crcr$#1$}}}

\title{
  Polyakov loop analysis with Dirac-mode expansion
}  

\ShortTitle{Polyakov loop analysis with Dirac-mode expansion}

\author{\speaker{Takumi Iritani}, Shinya Gongyo, and Hideo Suganuma \\
        Department of Physics, Kyoto University, Kitashirakawaoiwake, Sakyo,
        Kyoto 606-8502, Japan\\
        E-mail: \email{iritani@ruby.scphys.kyoto-u.ac.jp}
}

\abstract{
  In order to investigate the direct relation between
  confinement and chiral symmetry breaking in QCD,
  we investigate the Polyakov loop 
  in terms of the Dirac eigenmodes
  in both confined and deconfined phases.
  Using the Dirac-mode expansion method in SU(3) lattice QCD,
  we analyze the contribution of low-lying 
  and higher Dirac-modes to the Polyakov loop, respectively.
  In the confined phase below $T_c$,
  after removing low-lying Dirac-modes,
  the chiral condensate $\langle \bar{q} q\rangle$ is largely reduced,
  however, the Polyakov loop remains almost zero and $Z_3$-center symmetry is unbroken.
  These results indicate that the system is still in the confined phase
  without low-lying Dirac-modes.
  By higher Dirac-modes cut, the Polyakov loop also remains almost zero
  below $T_c$.
  We also analyze the Polyakov loop in the deconfined phase above $T_c$.
  We find that the Polyakov loop and
  $Z_3$-symmetry behavior 
  are insensitive to low-lying and higher Dirac-modes
  in both confined and deconfined phases.
}

\FullConference{
Xth Quark Confinement and the Hadron Spectrum,\\
October 8-12, 2012\\
TUM Campus Garching, Munich, Germany}

\begin{document}

\section{Introduction}

  Quantum chromodynamics (QCD)
  is the fundamental theory of the strong interaction,
  however, its non-perturbative properties
  such as confinement and chiral symmetry breaking
  are not yet well understood.
  In particular, to clarify the correspondence
  between confinement and chiral symmetry breaking
  is one of difficult and interesting subjects
  \cite{Gattringer:2006,Bruckmann:2007,Bilgici:2008,
  Synatschke:2008,Suganuma:2011,Gongyo:2012,Iritani:2012}.
  As an evidence of the close relation between them,
  lattice QCD calculation shows
  that simultaneous deconfinement and chiral phase transition
  at finite temperature \cite{Karsch:2002}.

  As shown in the Banks-Casher relation \cite{BanksCasher},
  the chiral condensate $\langle \bar{q}q \rangle$
  is proportional to the Dirac zero-mode density as
  \begin{equation}
    \langle \bar{q} q \rangle
    = - \lim_{m \rightarrow 0}
    \lim_{V\rightarrow \infty} \pi \langle \rho(0) \rangle,
  \end{equation}
  where $\rho(\lambda)$ is the Dirac spectral density.
  Thus, the low-lying Dirac eigenmodes
  directly relate to chiral symmetry breaking,
  however, their relation to confinement is still unclear.

  Therefore, it is interesting to analyze
  confinement in terms of the relevant degrees of freedom 
  for chiral symmetry breaking, i.e., the low-lying Dirac eigenmodes.
  For example, based on the Gattringer's formula \cite{Gattringer:2006},
  the Polyakov loop can be investigated
  by the sum of Dirac spectra with twisted boundary condition on lattice
  \cite{Bruckmann:2007,Bilgici:2008,Synatschke:2008}.
  In our previous studies \cite{Suganuma:2011,Gongyo:2012},
  we formulated the Dirac-mode expansion method in lattice QCD,
  and investigated the Dirac-mode dependence of 
  the Wilson loop and the interquark potential.
  It is also reported that
  hadrons still exist as bound states
  without ``chiral symmetry breaking'' by removing low-lying Dirac-modes
  \cite{Lang:2011,Glozman:2012}.

  In this paper,
  we investigate the Dirac-mode dependence
  of the Polyakov loop in both confined and deconfined phases
  at finite temperature, using Dirac-mode expansion method
  in lattice QCD \cite{Suganuma:2011,Gongyo:2012,Iritani:2012}.
  In Sec.~2, we review the Dirac-mode expansion method,
  and formulate the Dirac-mode projected Polyakov loop.
  In Sec.~3, we perform the lattice QCD calculations for
  the Polyakov loop with Dirac-mode projection.
  Section 4 is devoted for the summary.

\section{Dirac-mode expansion method in lattice QCD}
  Here, we introduce the Dirac-mode expansion technique in lattice QCD
  \cite{Suganuma:2011,Gongyo:2012,Iritani:2012},
  and formulation of the Dirac-mode projected Polyakov loop.

  \subsection{Dirac-mode expansion in lattice QCD}
  Using the link-variable $U_\mu \in \mathrm{SU}(N_c)$,
  the Dirac operator $\Slash{D} = \gamma_\mu D_\mu$ is given by
  \begin{equation}
    \Slash{D}_{x,y} \equiv \frac{1}{2a}
    \sum_{\mu = 1}^4
    \gamma_\mu \left[ U_\mu(x) \delta_{x+\hat{\mu},y}
    - U_{-\mu}(x) \delta_{x-\hat{\mu},y}\right],
    \label{eq:dirac-naive}
  \end{equation}
  with a lattice spacing $a$, and $U_{-\mu}(x) \equiv U_\mu^\dagger(x-\hat{\mu})$.
  Here, $\gamma$-matrix is defined to be hermitian, i.e.,
  $\gamma_\mu^\dagger = \gamma_\mu$.
  Thus, $\Slash{D}$ becomes an antihermitian operator,
  and Dirac eigenvalues are pure imaginary.
  We introduce the normalized Dirac eigenstate $|n \rangle$, which satisfies
  \begin{equation}
    \Slash{D}|n\rangle = i\lambda_n | n \rangle,
  \end{equation}
  with $\lambda_n \in \mathbf{R}$,
  and an eigenfunction $\psi_n(x)$ is expressed as
  \begin{equation}
    \psi_n(x) \equiv \langle x | n \rangle,
  \end{equation}
  which satisfies $\Slash{D}\psi_n = i \lambda_n \psi_n$.

  We introduce the operator formalism in lattice QCD 
  \cite{Suganuma:2011,Gongyo:2012,Iritani:2012},
  which is constructed from the link-variable operator $\hat{U}_\mu$.
  The link-variable operator is defined by 
  the matrix element as
  \begin{equation}
    \langle x | \hat{U}_\mu | y \rangle = U_\mu(x) \delta_{x+\hat{\mu},y},
  \end{equation}
  using the original link-variable $U_\mu(x)$.
  We define the Dirac-mode matrix element $\langle  n| \hat{U}_\mu | m \rangle$ as
  \begin{eqnarray}
    \langle n | \hat{U}_\mu | m \rangle
    &=& \sum_x \langle n | x \rangle \langle x |
    \hat{U}_\mu | x + \hat{\mu} \rangle
    \langle x + \hat{\mu} | m \rangle
    = \sum_x \psi_n^\dagger(x) U_\mu(x) \psi_m(x+\hat{\mu}),
  \end{eqnarray}
  with the Dirac eigenfunction $\psi_n(x)$.

  Considering the completeness relation $\sum_n | n\rangle \langle n| = 1$,
  any operator $\hat{O}$ can be expressed as
  \begin{equation}
    \hat{O} = \sum_n \sum_m | n \rangle \langle n | \hat{O} | m \rangle
    \langle m |,
    \label{eq:operator-expansion}
  \end{equation}
  using the Dirac-mode basis.
  Note that this procedure is just
  the insertion of unity, and it is mathematically correct.
  This expansion (\ref{eq:operator-expansion})
  is the mathematical basis of the Dirac-mode expansion method 
  \cite{Suganuma:2011,Gongyo:2012,Iritani:2012}.
  Next, we consider the Dirac-mode projection by introducing
  projection operator as
  \begin{equation}
    \hat{P} \equiv \sum_{n \in \mathcal{A}} | n \rangle \langle n |,
  \end{equation}
  for arbitrary set $\mathcal{A}$ of eigenmodes.
  For example, IR and UV mode-cut operators are given by
  \begin{equation}
    \hat{P}_{\rm IR} \equiv \sum_{|\lambda_n| \geq \Lambda_{\rm IR}} | n \rangle \langle n |, 
    \qquad
    \hat{P}_{\rm UV} \equiv \sum_{|\lambda_n| \leq \Lambda_{\rm UV}} | n \rangle \langle n |,
  \end{equation}
  with the IR/UV cutoff scale $\Lambda_{\rm IR}$ and $\Lambda_{\rm UV}$.

  Using the projection operator $\hat{P}$,
  the Dirac-mode projected link-variable operator
  is given by
  \begin{equation}
    \hat{U}_\mu^P \equiv \hat{P} \hat{U}_\mu \hat{P} 
    = \sum_{n \in \mathcal{A}} \sum_{m \in \mathcal{A}} 
    | n \rangle \langle n | \hat{U}_\mu | m \rangle \langle m |.
  \end{equation}
  We can investigate the Dirac-mode dependence
  of various kinds of quantities, e.g., the Wilson loop \cite{Suganuma:2011,Gongyo:2012},
  using the projected link-variable $\hat{U}_\mu^P$ instead of the original
  link-variable operator $\hat{U}_\mu$.

  \subsection{Polyakov loop operator and Dirac-mode projection}
  Next, we formulate the Dirac-mode projected Polyakov loop.
  Here, we consider  the periodic
  SU(3) lattice with 
  the space-time volume $V = L^3 \times N_t$
  and the lattice spacing $a$.
  In the operator formalism of lattice QCD,
  the Polyakov loop operator is given by
  \begin{equation}
    \hat{L}_P \equiv \frac{1}{3V} \prod_{i=1}^{N_t} \hat{U}_4
    = \frac{1}{3V} \hat{U}_4^{N_t}
  \end{equation}
  with the temporal link-variable operator $\hat{U}_4$.
  By the functional trace ``Tr'',
  the Polyakov loop operator coincides with the standard definition as
  \begin{eqnarray}
    \mathrm{Tr} \ \hat{L}_P 
    &=& \frac{1}{3V} \mathrm{Tr} \ 
    \large\{ \prod_{i=1}^{N_t} \hat{U}_4 \large\} = \frac{1}{3V} \mathrm{tr}
    \sum_{\vec{x},t} 
    \langle \vec{x},t | \prod_{i=1}^{N_t} \hat{U}_4 | \vec{x},t \rangle \nonumber  \\
    &=& \frac{1}{3V} \mathrm{tr} 
    \sum_{\vec{x},t} \langle \vec{x},t | \hat{U}_4 | \vec{x},t + a \rangle 
      \langle \vec{x},t+a | \hat{U}_4 | \vec{x},t+2a \rangle 
      \cdots \langle \vec{x},t + (N_t-1)a | \hat{U}_4 | \vec{x},t \rangle \nonumber \\
    &=& \frac{1}{3V} \mathrm{tr}
      \sum_{\vec{x},t} U_4(\vec{x},t) U_4(\vec{x},t+a) \cdots U_4(\vec{x},t+(N_t-1)a)
      = \langle L_P \rangle,
  \end{eqnarray}
  where ``tr'' denotes the trace over SU(3) color index.

  We define the Dirac-mode projected Polyakov loop $\langle L_P^{\rm proj.} \rangle$ as
  \begin{eqnarray}
    L_{P}^{\rm proj.} 
    &\equiv& \frac{1}{3V} \mathrm{Tr} \ \large\{ \prod_{i=1}^{N_t} \hat{U}_4^P \large\} = 
    \frac{1}{3V} \mathrm{Tr} \left\{
    \hat{P}\hat{U}_4 \hat{P} \hat{U}_4 \hat{P} \cdots 
    \hat{P} \hat{U}_4 \hat{P} \right\} \nonumber \\
    &=& \frac{1}{3V}\mathrm{tr} \sum_{n_1,n_2,\dots, n_{N_t} \in \mathcal{A}}
      \langle n_1 | \hat{U}_4 | n_2 \rangle \langle n_2 | \hat{U}_4 | n_3 \rangle
      \cdots \langle n_{N_t} | \hat{U}_4 | n_1 \rangle.
    \label{eq:Lproj}
  \end{eqnarray}
  In particular, the IR and the UV Dirac-mode projected Polyakov loop 
  are denoted as
  \begin{eqnarray}
    \langle L_P \rangle_{\rm IR}
    &\equiv& \frac{1}{3V}  \mathrm{tr}
    \sum_{|\lambda_{n_i}| \geq \Lambda_{\rm IR}}
    \langle n_1 | \hat{U}_4 | n_2 \rangle
    \cdots \langle n_{N_t} | \hat{U}_4 | n_1 \rangle, \\
    \langle L_P \rangle_{\rm UV}
    &\equiv& \frac{1}{3V} \mathrm{tr}
    \sum_{|\lambda_{n_i}| \leq \Lambda_{\rm UV}}
    \langle n_1 | \hat{U}_4 | n_2 \rangle
    \cdots \langle n_{N_t} | \hat{U}_4 | n_1 \rangle,
  \end{eqnarray}
  with the IR/UV  eigenvalue cutoff $\Lambda_{\rm IR}$ and $\Lambda_{\rm UV}$.

\section{Lattice QCD calculation for Dirac-mode projected Polyakov loop}
  In this section,
  we calculate the Dirac-mode projected Polyakov loop
  using SU(3) lattice QCD at the quenched level.
  We evaluate the full Dirac eigenmodes using LAPACK \cite{LAPACK}.
  For actual calculation,
  we use the eigenmode basis of the Kogut-Susskind (KS) operator of
  \begin{equation}
    D_{x,y}^{\rm KS} \equiv \frac{1}{2a}
    \sum_{\mu=1}^4 \eta_\mu(x) \left[ U_\mu(x) \delta_{x+\hat{\mu},y}
    - U_{-\mu}(x) \delta_{x-\hat{\mu},y}\right],
  \end{equation}
  with 
    $\eta_1(x) \equiv 1$ and $\eta_\mu(x) \equiv (-1)^{x_1 +\cdots + x_{\mu-1}}$
    ($\mu \geq 2$)
    in order to reduce the computational cost.
  The use of the KS-Dirac operator gives the same result as the original Dirac operator
  in Eq.~(\ref{eq:dirac-naive})
  for the Polyakov loop.

\subsection{The confined phase}

  First, we analyze the Polyakov loop in the confined phase below $T_c$.
  Here, we use $6^4$ lattice with $\beta = 5.6$,
  which corresponds to the lattice spacing $a \simeq 0.25$~fm 
  and $T \equiv 1/(N_ta) \simeq 0.13$~GeV
  \cite{Suganuma:2011,Gongyo:2012,Iritani:2012}.

\begin{figure}
  \centering
  \includegraphics[width=0.325\textwidth,clip]{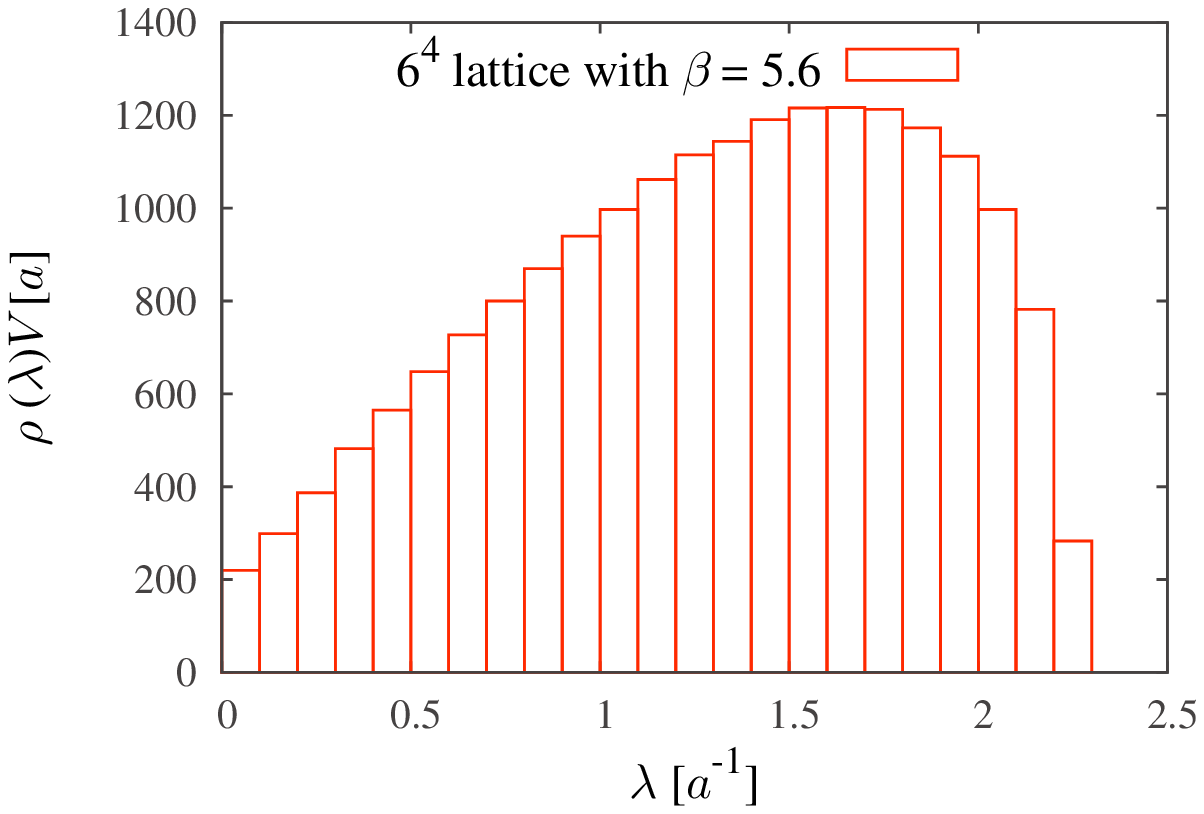}
  \includegraphics[width=0.325\textwidth,clip]{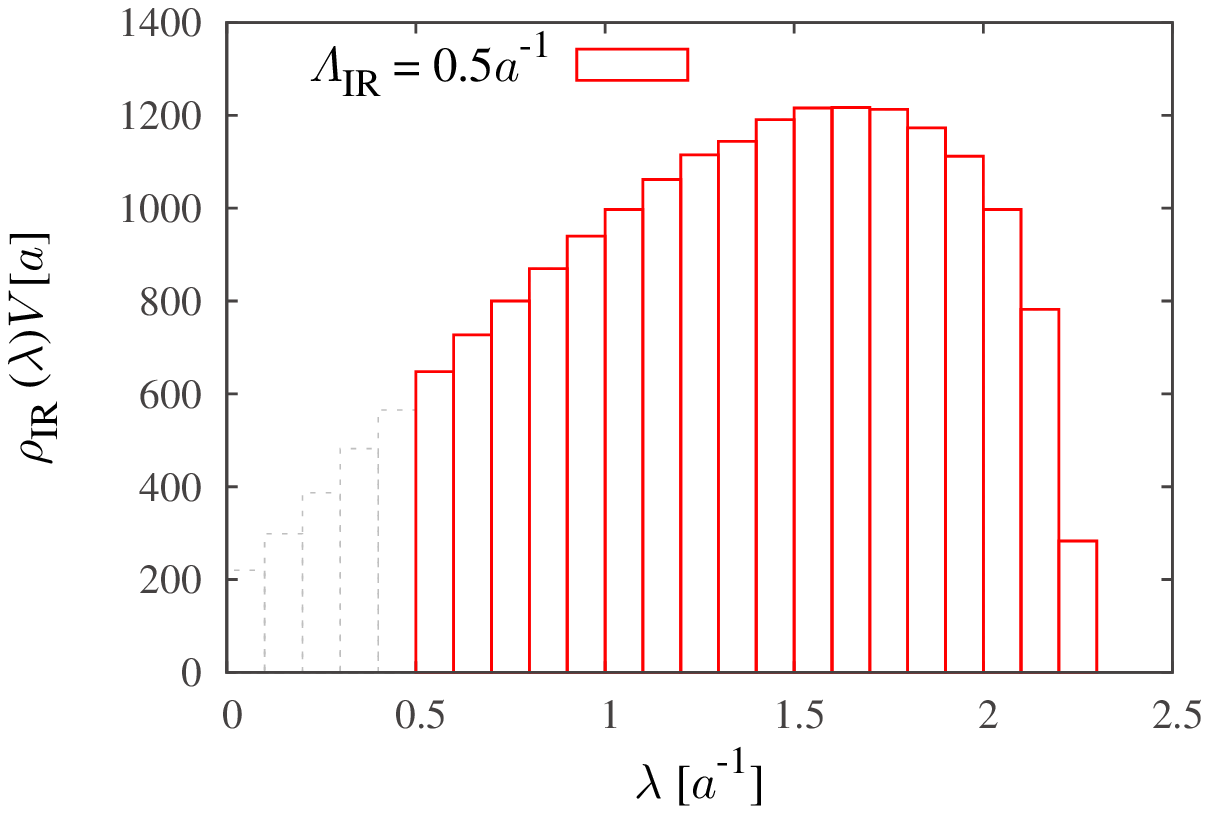}
  \includegraphics[width=0.325\textwidth,clip]{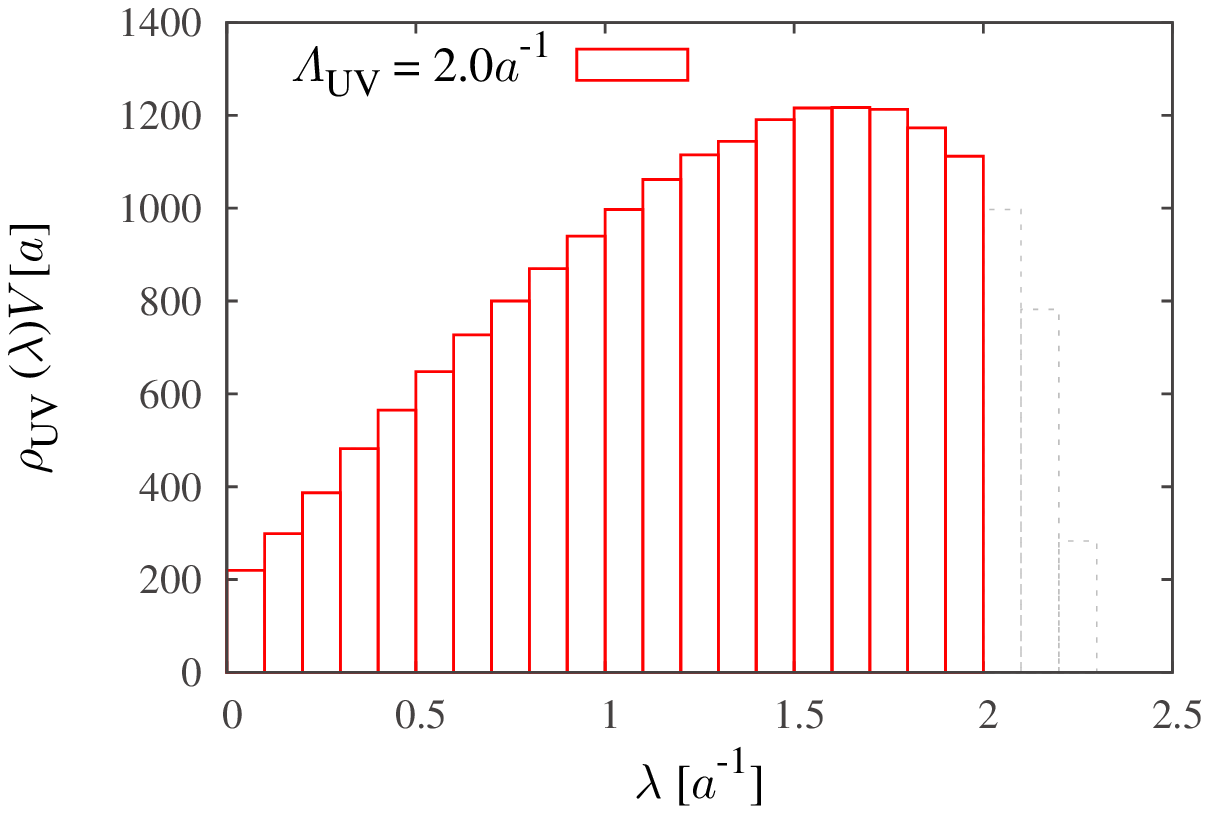}
  \caption{\label{fig:DiracSpectrumCut} 
    The spectral density $\rho(\lambda)$ of the Dirac operator
    on $6^4$ lattice with $\beta = 5.6$, i.e., $a = 0.25$~fm:
    (a) original spectral density,
    (b) IR-cut $\rho_{\rm IR}(\lambda)$ at $\Lambda_{\rm IR} = 0.5a^{-1}$,
    (c) UV-cut $\rho_{\rm UV}(\lambda)$ at $\Lambda_{\rm UV} = 2.0a^{-1}$.
  }
\end{figure}

  Figure \ref{fig:DiracSpectrumCut} shows
  the lattice QCD results for the Dirac spectral density $\rho(\lambda)$,
  and IR/UV-cut spectral density
    $\rho_{\rm IR}(\lambda) \equiv \rho(\lambda)\theta(|\lambda|-
    \Lambda_{\rm IR})$, 
    $\rho_{\rm UV}(\lambda) \equiv \rho(\lambda)
    \theta( \Lambda_{\rm UV} - | \lambda | )$
  with $\Lambda_{\rm IR} = 0.5a^{-1}$ and $\Lambda_{\rm UV} = 2.0a^{-1}$.
  The total number of the KS Dirac-mode is $L^3 \times N_t \times 3 = 3888$,
  and both mode cuts correspond to removing about 400 modes.

\begin{figure}
  \centering
  \includegraphics[width=0.325\textwidth,clip]{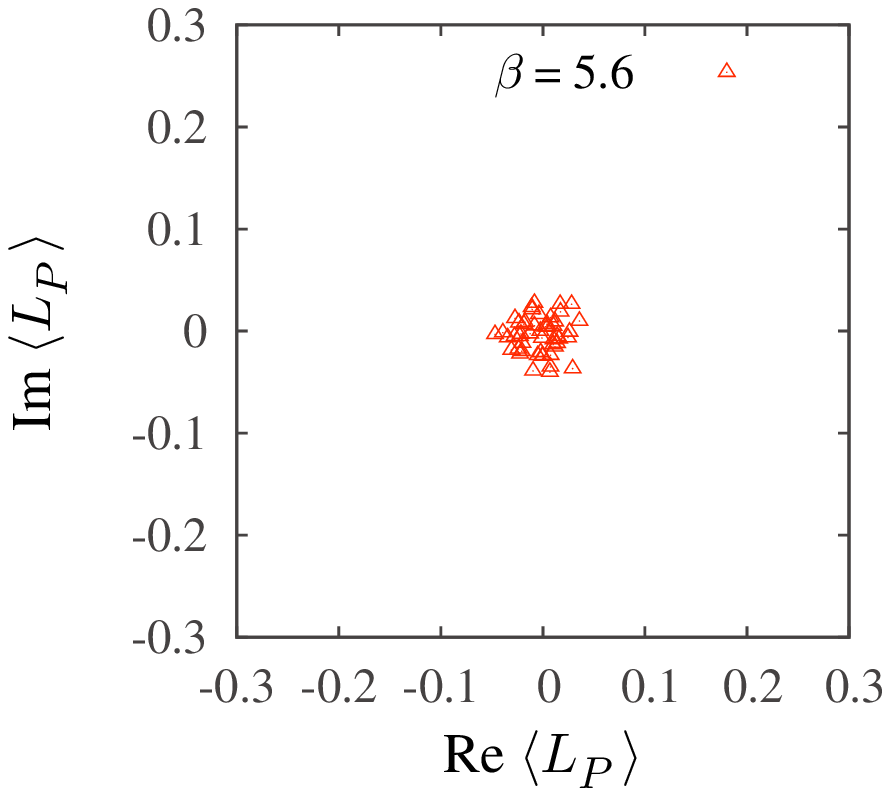}
  \includegraphics[width=0.325\textwidth,clip]{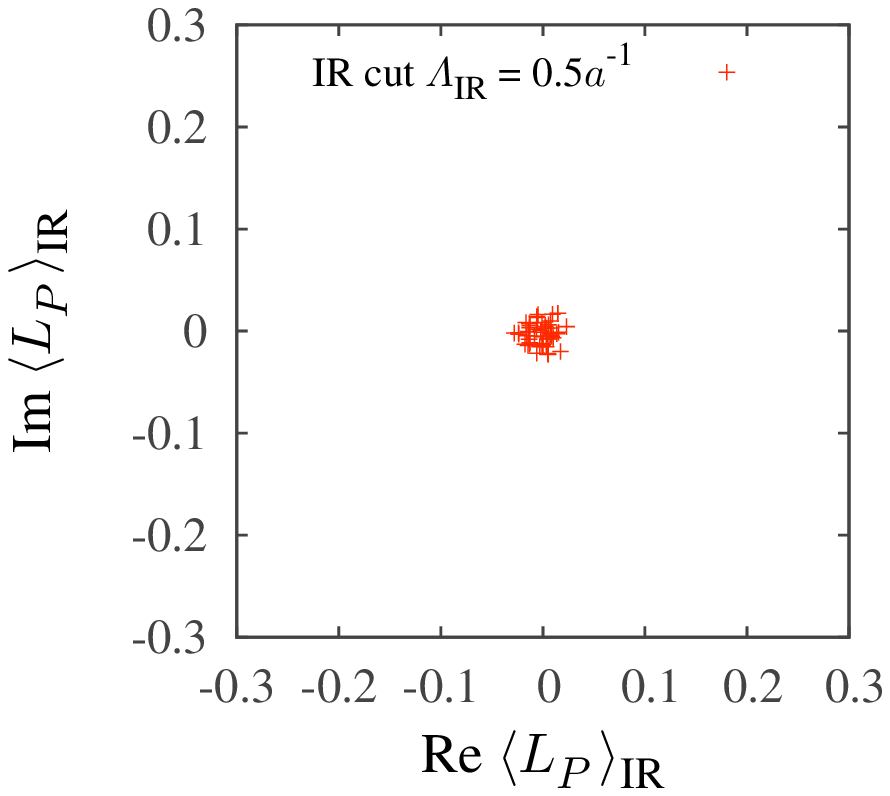}
  \includegraphics[width=0.325\textwidth,clip]{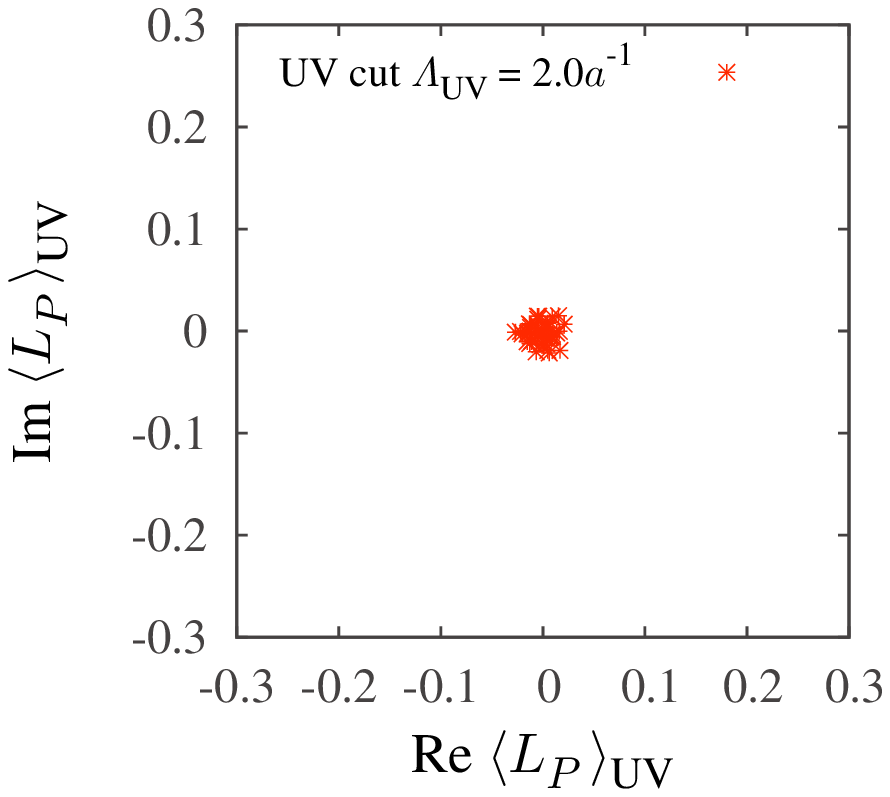}
  \caption{\label{fig:PolyakovScatterConfined} 
    The scatter plot of the Polyakov loop in the confined phase
    on $6^4$ lattice with $\beta = 5.6$, i.e.,
    $a = 0.25$~fm and $T \equiv 1/(N_ta) \simeq 0.2$~GeV.
    (a) The original (no Dirac-mode cut) Polyakov loop.
    (b) The low-lying Dirac-mode cut at $\Lambda_{\rm IR} = 0.5a^{-1}$.
    (c) The higher Dirac-mode cut at $\Lambda_{\rm UV} = 2.0a^{-1}$.
  }
\end{figure}

  Figure \ref{fig:PolyakovScatterConfined}
  shows the scatter plot of the original Polyakov loop,
  low-lying Dirac-modes cut at $\Lambda_{\rm IR} = 0.5a^{-1}$,
  and the higher Dirac-modes cut at $\Lambda_{\rm UV} = 2.0a^{-1}$, respectively.
  As shown in Fig.~\ref{fig:PolyakovScatterConfined}(a),
  the Polyakov loop is almost zero,
  i.e., $\langle L_P \rangle \simeq 0$,
  which indicates the confined phase.

  Then, we consider low-lying Dirac-modes projection,
  which leads to the effective restoration
  of chiral symmetry breaking
  \cite{Gongyo:2012,BanksCasher,Lang:2011,Glozman:2012}.
  In the presence of the IR cut $\Lambda_{\rm IR}$,
  the quark condensate is given by
  \begin{equation}
    \langle \bar{q}q \rangle_{\rm IR} = - \frac{1}{V}
    \sum_{\lambda_n \geq \Lambda_{\rm IR}} \frac{2m}{\lambda_n^2+m^2}.
    \label{eq:qbarq-ir-cut}
  \end{equation}
  At the IR cut parameter $\Lambda_{\rm IR} = 0.5a^{-1} \simeq 0.4$~GeV,
  only 2\% of the quark condensate remains as 
  $\langle \bar{q}q \rangle_{\rm IR}/\langle \bar{q}q \rangle \simeq 0.02$
  around the physical region $m \simeq 5$~MeV \cite{Gongyo:2012}.
  However, as shown in Fig.~\ref{fig:PolyakovScatterConfined}(b),
  the Polyakov loop $\langle L_P \rangle_{\rm IR}$ remains almost zero
  and $Z_3$-center symmetry is unbroken,
  and these facts indicate that the system still remains in the confined phase,
  even without chiral symmetry breaking.

  In addition to the low-lying mode cut,
  we show the higher Dirac-modes cut in Fig.~\ref{fig:PolyakovScatterConfined}(c).
  In this case, the chiral condensation is almost unchanged,
  and the Polyakov loop remains almost zero.
  Therefore, the Polyakov loop is insensitive
  to both low-lying and higher Dirac eigenmodes.

\subsection{The deconfined phase at high temperature}
  Next, we investigate the Polyakov loop 
  in the deconfined phase at high temperature.
  Here, we use $6^3 \times 4$ lattice with $\beta = 6.0$,
  which corresponds to $a = 0.10$~fm and
  $T \equiv 1/(N_ta) \simeq 0.5$~GeV.
  The total number of the KS Dirac-mode
  is $L^3 \times N_t \times 3 = 2592$.

\begin{figure}
    \centering
    \includegraphics[width=0.325\textwidth,clip]{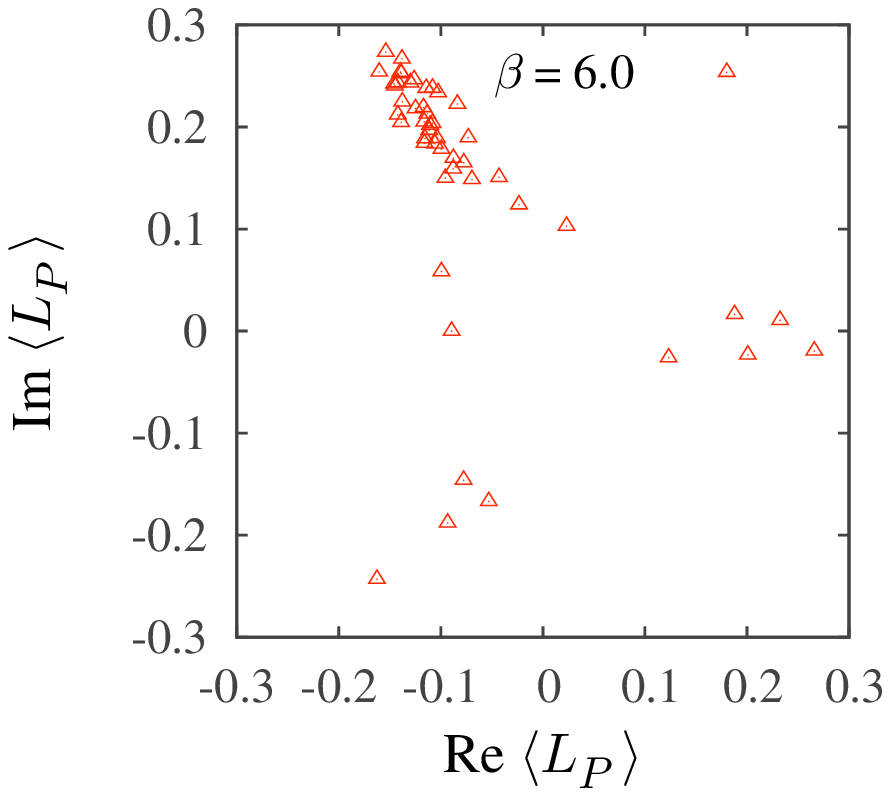}
    \includegraphics[width=0.325\textwidth,clip]{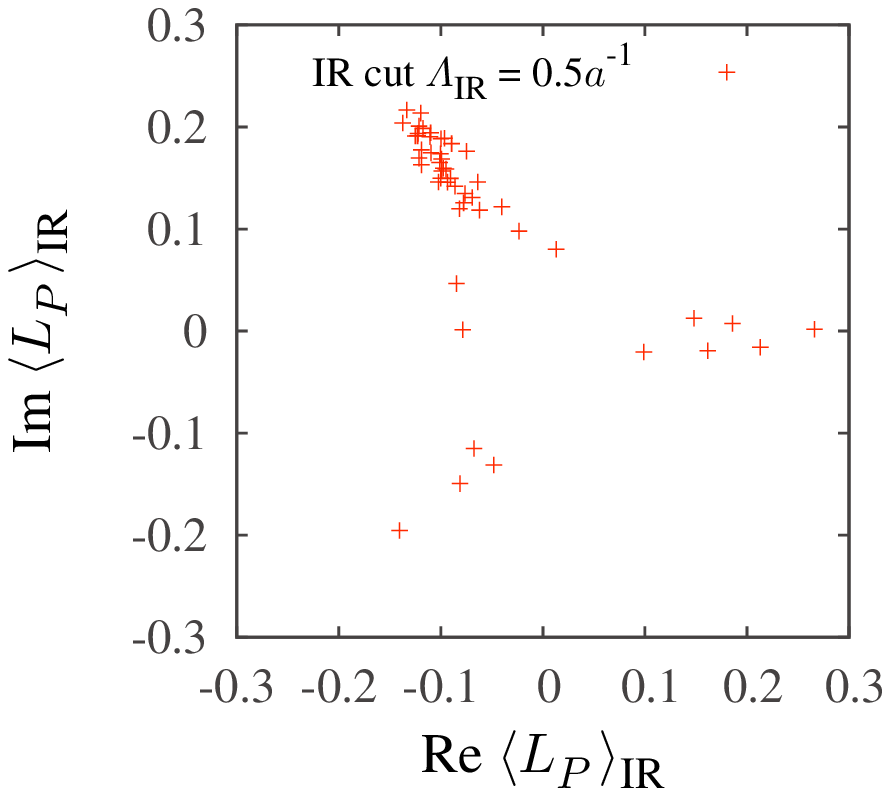}
    \includegraphics[width=0.325\textwidth,clip]{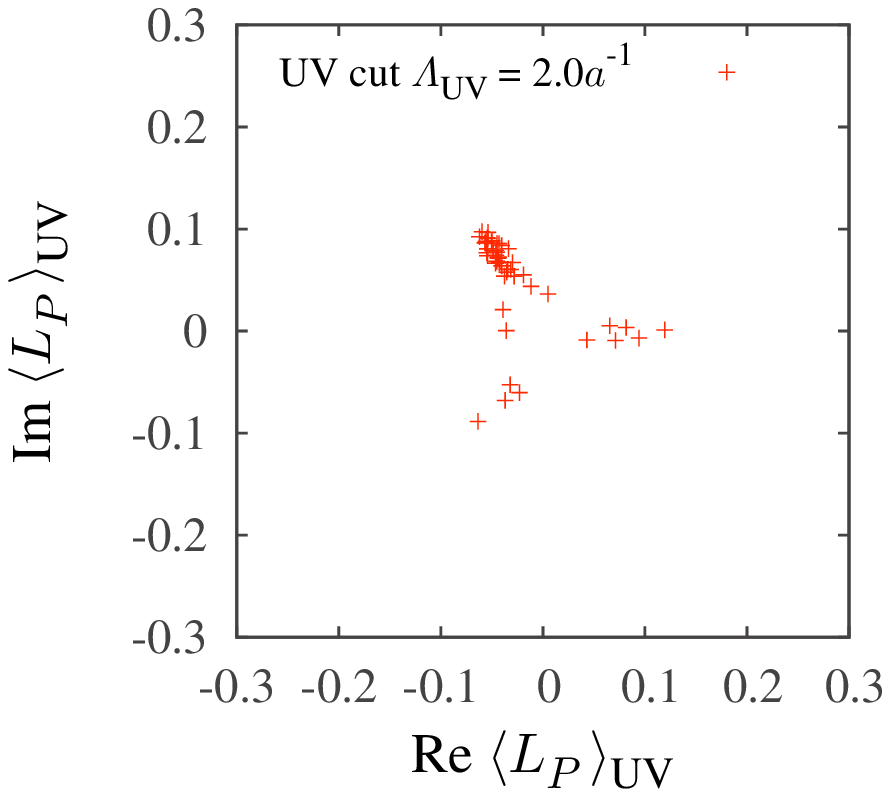}
    \caption{\label{fig:PolyakovScatterDeconfined}
      The scatter plot of the Polyakov loop in the deconfined phase
      on $6^3 \times 4$ lattice with $\beta = 6.0$, 
      i.e., $a = 0.10$~fm and $T \equiv 1/(N_ta) \simeq 0.5$~GeV.
      (a) The original (no Dirac-mode cut) Polyakov loop.
      (b) The low-lying Dirac-mode cut at $\Lambda_{\rm IR} = 0.5a^{-1}$.
      (c) The higher Dirac-mode cut at $\Lambda_{\rm UV} = 2.0a^{-1}$.
    }
  \end{figure}

  We show the original Polyakov loop, 
  typical low-lying mode cut at $\Lambda_{\rm IR} = 0.5a^{-1}$,
  and higher mode cut at $\Lambda_{\rm UV} = 2.0a^{-1}$
  in Fig.~\ref{fig:PolyakovScatterDeconfined}.
  These mode cuts correspond to removing about 200 eigenmodes.
  As shown in Fig.~\ref{fig:PolyakovScatterDeconfined}(a),
  the Polyakov loop has non-zero expectation value $\langle L_P \rangle \neq 0$,
  and shows the center group $Z_3$ structure on the complex plane.
  These behaviors indicate the deconfined phase.

  As shown in Figs.~\ref{fig:PolyakovScatterDeconfined}(b) and (c),
  the Polyakov loop shows the characteristic behaviors 
  in the deconfined phase even after removing low-lying or higher Dirac-modes.
  To be strict,
  the UV-cut Polyakov loop has a smaller absolute value than the IR-cut Polyakov loop,
  although the number of UV-cut modes is comparable to that of the IR-cut case.
  This suggests that contributions of
  the higher Dirac-modes are much larger than low-lying modes
  \cite{Bruckmann:2007}.
  However, apart from the normalization,
  both IR and UV cut Polyakov loops show
  the characteristic $Z_3$-pattern in the deconfined phase,
  and hence these Dirac-modes seem to be insensitive to the Polyakov loop properties.

\begin{figure}
    \centering
    \includegraphics[width=0.325\textwidth,clip]{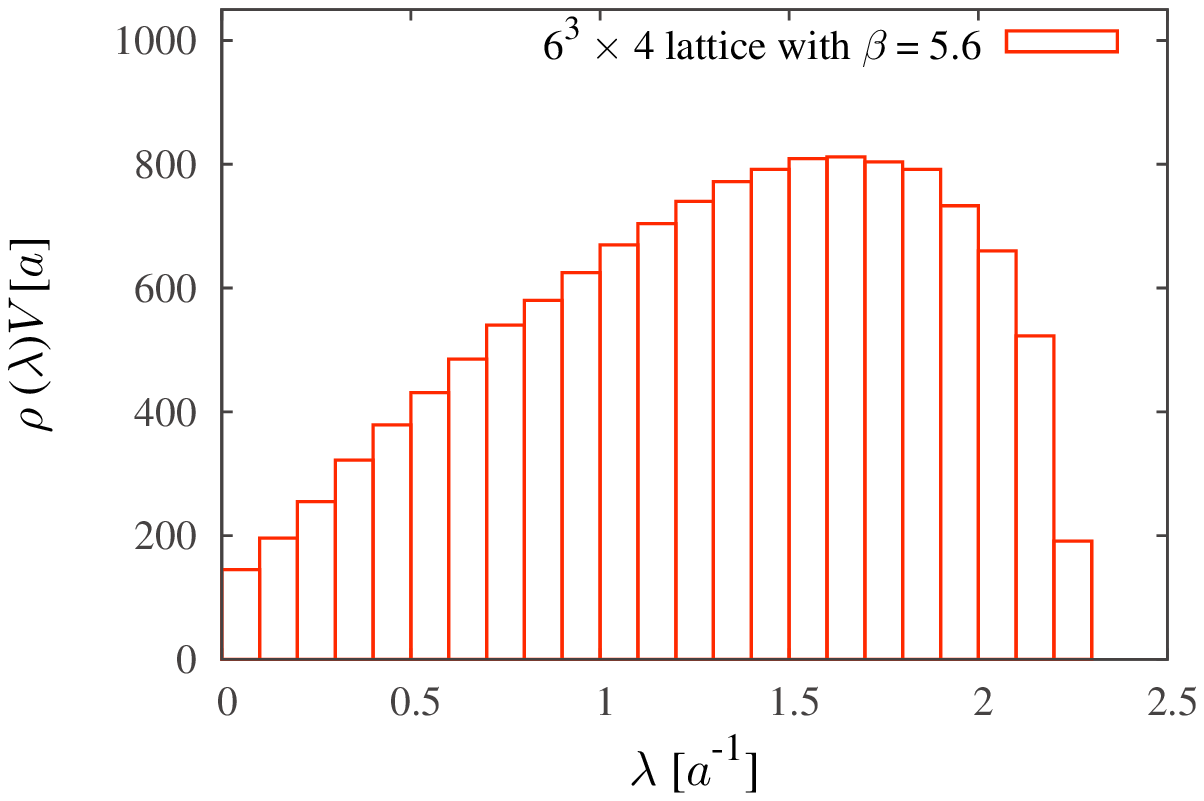}
    \includegraphics[width=0.325\textwidth,clip]{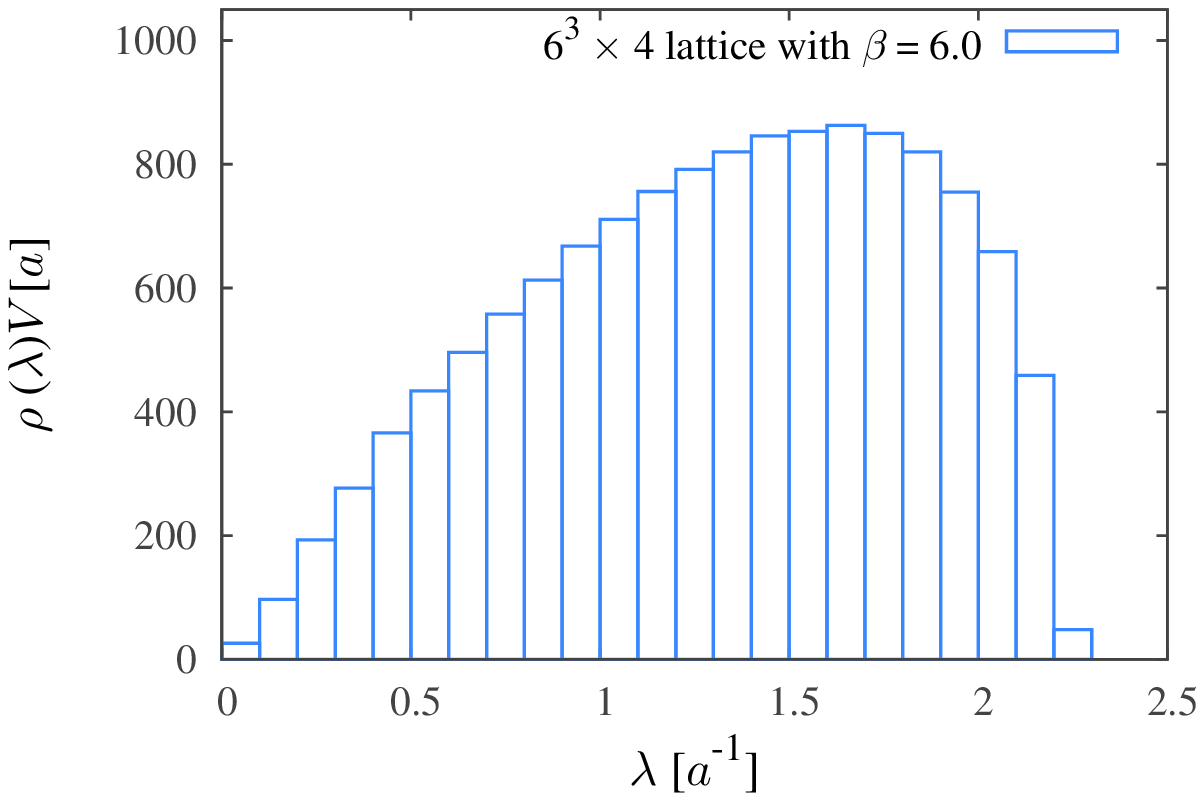}
    \includegraphics[width=0.325\textwidth,clip]{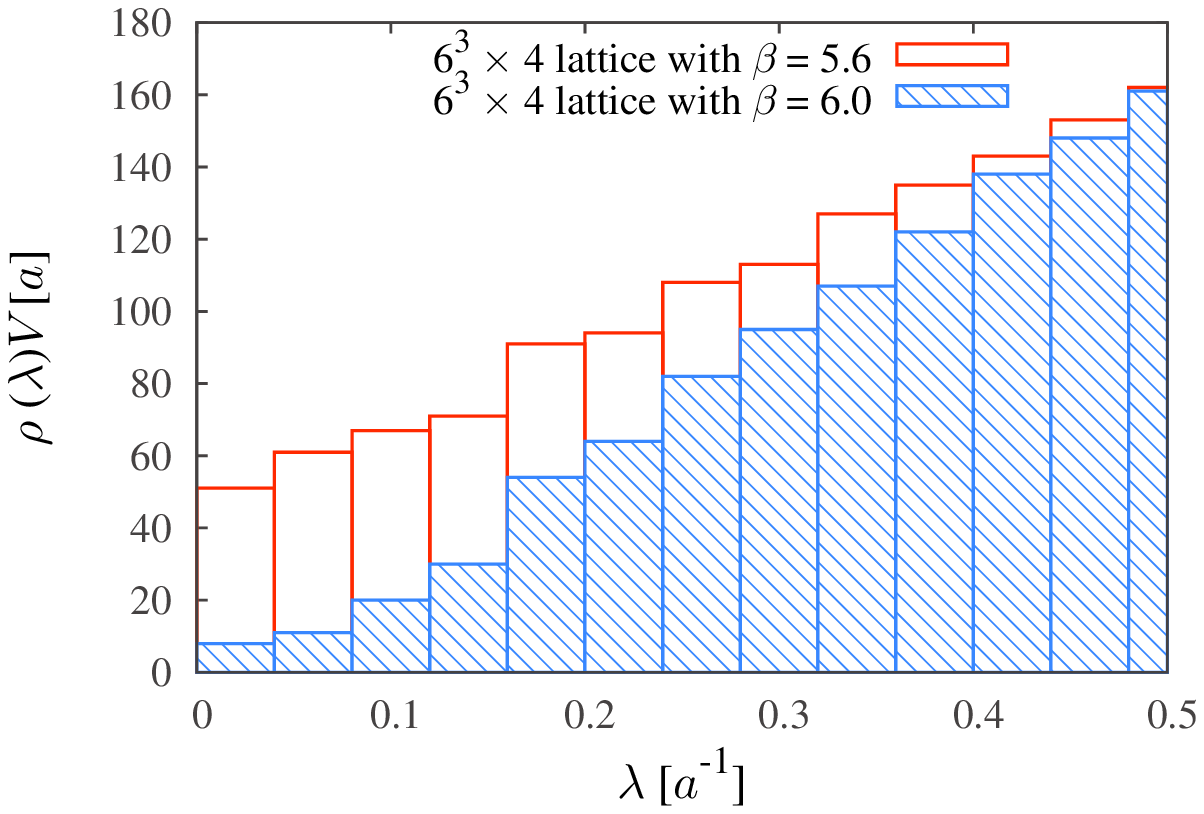}
    \caption{ \label{fig:DiracCompare}
      The Dirac spectral densities in confined and deconfined phases, respectively
      (a) $6^3 \times 4$ lattice with $\beta = 5.6$ in the confined phase. 
      (b) $6^3 \times 4$ lattice with $\beta = 6.0$ in the deconfined phase.
      (c) The comparison between confined and deconfined phases
      on low-lying spectral densities.
    }
  \end{figure}

  Figure~\ref{fig:DiracCompare} shows
  the Dirac spectral densities in confined and deconfined phases
  on $6^3 \times 4$ lattice with $\beta = 5.6$ and $6.0$, respectively.
  We also compare their low-lying spectral densities in Fig.~\ref{fig:DiracCompare}(c).
  In the deconfinement phase, the low-lying Dirac-modes are suppressed,
  which leads to the chiral restoration.
  The chiral condensate is also reduced by the IR cut 
  of the Dirac-modes as in Eq.~(\ref{eq:qbarq-ir-cut}).
  On the other hand, 
  there seems no clear correspondence 
  between the Dirac spectral densities and the Polyakov loop.

\subsection{$\beta$-dependence of the Dirac-mode projected Polyakov loop}
  Finally, we investigate $\beta$-dependence of the 
  Dirac-mode projected Polyakov loop.
  Here, we adopt $6^3 \times 4$ lattice with $\beta = 5.4 \sim 6.0$.

\begin{figure}
  \centering
  \includegraphics[width=0.49\textwidth,clip]{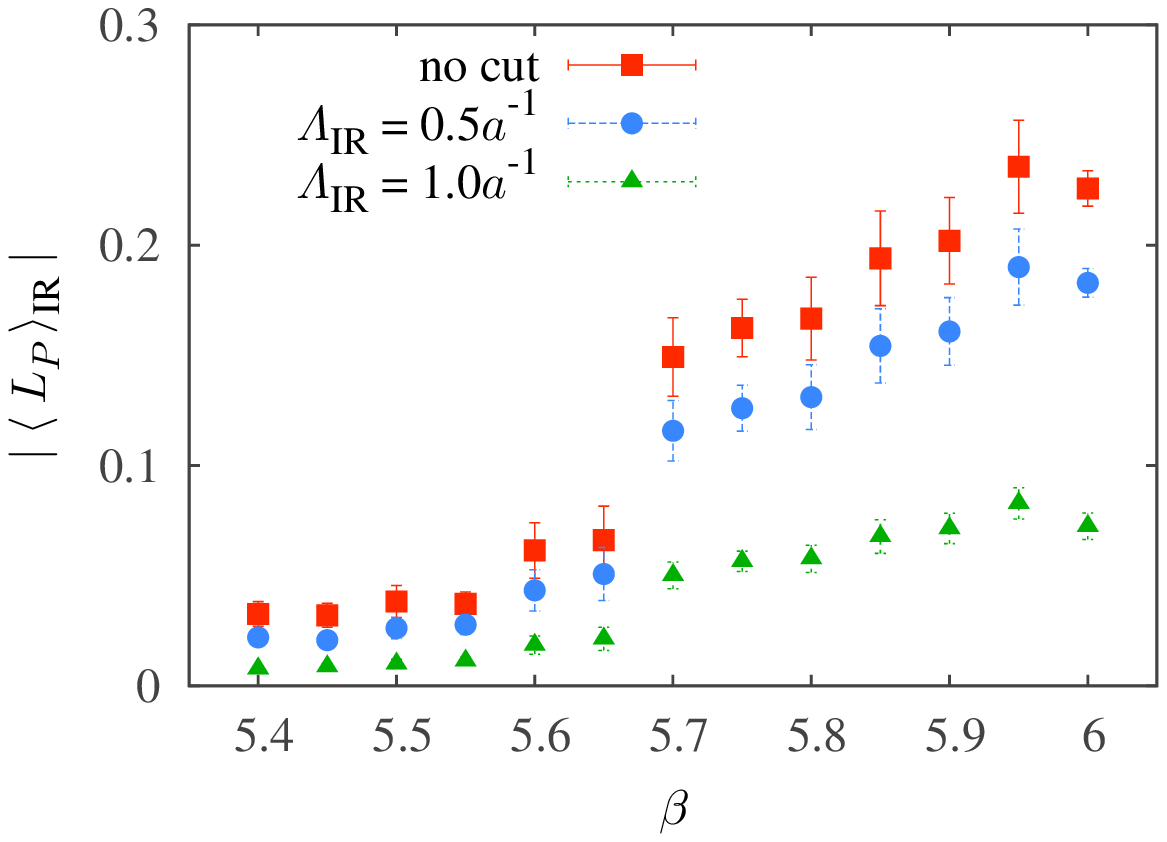}
  \includegraphics[width=0.49\textwidth,clip]{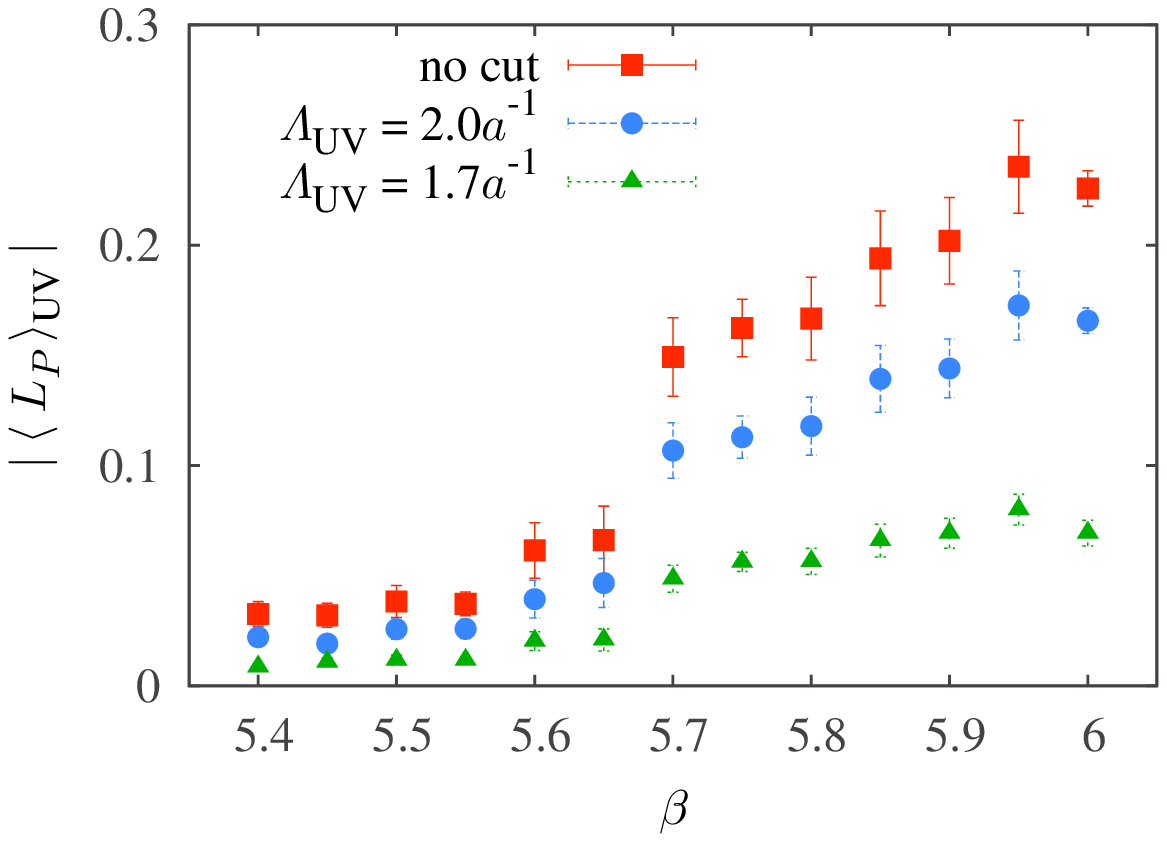}
  \caption{ \label{fig:PolyakovDiracCut}
    $\beta$-dependence of the absolute value of the Polyakov loop
    on $6^3 \times 4$ lattice.
    (a) The IR Dirac-mode cut with $\Lambda_{\rm IR} = 0.5a^{-1}$ and $1.0a^{-1}$.
    (b) The UV Dirac-mode cut with $\Lambda_{\rm UV} = 2.0a^{-1}$ and $1.7a^{-1}$.
  }
\end{figure}

  Figure \ref{fig:PolyakovDiracCut}
  is the absolute values of the Polyakov loop with typical Dirac-mode projections,
  and the original Polyakov loop data are also added for comparison.
  In this lattice volume, 
  the deconfinement 
  phase transition occurs around $\beta = 5.6 \sim 5.7$.
  As shown in Fig.~\ref{fig:PolyakovDiracCut},
  both low-lying and higher Dirac-mode projected Polyakov loops
  show the similar $\beta$-dependence as the original data,
  apart from a normalization factor.
  This Dirac-mode insensitivity of the Polyakov loop
  is consistent with the results in the previous subsections.

\section{Summary}
  In this paper, we have analyzed the Polyakov loop 
  in terms of the Dirac eigenmodes using SU(3) lattice QCD.
  We have carefully removed 
  relevant degrees of freedom for chiral symmetry breaking
  from the Polyakov loop.

  In the confined phase below $T_c$, 
  the Polyakov loop is almost zero, i.e., $\langle L_P \rangle \simeq 0$.
  By removing low-lying Dirac-modes,
  the chiral condensate $\langle \bar{q} q \rangle$ is largely reduced.
  However, we have found that the Polyakov loop remains almost zero
  as $\langle L_P \rangle_{\rm IR} \simeq 0$
  even without low-lying Dirac-modes,
  and this fact indicates that the system still remains in the confined phase.
  We have also investigated contributions from 
  higher Dirac-modes to the Polyakov loop,
  and have found no change of the Polyakov loop without higher Dirac-modes.
  Therefore, there seems no specific region of the Dirac eigenmodes
  essential for the Polyakov loop.

  We have also investigated the Polyakov loop 
  in the deconfined phase at high temperature.
  In the deconfined phase, the Polyakov loop has non-zero expectation value,
  i.e., $\langle L_P \rangle \neq 0$,
  which distributes around $Z_3$ elements in the complex plane.
  These characteristic behaviors
  also remain without low-lying and higher Dirac eigenmodes.
  Therefore, the Polyakov loop and the $Z_3$-symmetry behavior
  do not depend on low-lying and higher Dirac eigenmodes
  in both confinement and deconfinement phases.

  Here, we comment on the related studies about
  the correspondence between the Dirac eigenmodes and confinement.
  In the previous studies \cite{Suganuma:2011,Gongyo:2012},
  we investigated Dirac-mode dependence of the Wilson loop,
  and found that the confining potential 
  survives without low-lying Dirac eigenmodes.
  The Graz group also reported that 
  hadrons still remain as bound states
  without chiral symmetry breaking by removing low-lying Dirac-modes
  \cite{Lang:2011,Glozman:2012},
  which seems to suggest the existence of the confining force.

  These lattice QCD studies suggest
  that there is no direct relation between
  chiral symmetry breaking and confinement
  through the Dirac eigenmodes.
  For further investigation of correspondence between these phenomena,
  it is also interesting to analyze
  chiral symmetry breaking from
  the relevant eigenmodes of confinement \cite{Iritani:2012fp}.

\section*{Acknowledgements}
  The lattice QCD calculations have been done 
  on NEC-SX8 and NEC-SX9 at Osaka University. 
  This work is in part supported by 
  a Grant-in-Aid for JSPS Fellows [No.23-752, 24-1458] 
  and the Grant for Scientific Research 
  [(C) No.23540306, Priority Areas ``New Hadrons'' (E01:21105006)] 
  from the Ministry of Education, Culture, Science and Technology (MEXT) of Japan.

\end{document}